\documentclass[conference]{IEEEtran}
\usepackage{url}
\usepackage{cite}
\usepackage{amsmath,amssymb,amsfonts}
\usepackage{algorithmic}
\usepackage{graphicx}
\usepackage{textcomp}
\usepackage{epsfig}
\usepackage{epsf}
\usepackage{bmpsize}
\usepackage{boxedminipage}
\usepackage[table]{xcolor}
\usepackage{ulem}
\usepackage{ulem}
\usepackage{textcomp} 
\newcommand\BlackCell[1]{%
  \multicolumn{1}{c|}{\cellcolor{black}\textcolor{white}{#1}}
}

\usepackage{placeins}
\usepackage{listings}
\usepackage{lipsum}
\definecolor{codegreen}{rgb}{0,0.6,0}
\definecolor{codegray}{rgb}{0.5,0.5,0.5}
\definecolor{codepurple}{rgb}{0.58,0,0.82}
\definecolor{backcolour}{rgb}{0.95,0.95,0.92}

\lstdefinestyle{mystyle}{
  backgroundcolor=\color{white}, %backcolour},   commentstyle=\color{codegreen},
  keywordstyle=\color{magenta},
  numberstyle=\tiny\color{codegray},
  stringstyle=\color{codepurple},
  basicstyle=\ttfamily\footnotesize,
  breakatwhitespace=false,         
  breaklines=true,                 
  captionpos=b,                    
  keepspaces=true,                 
  numbers=left,                    
  numbersep=5pt,                  
  showspaces=false,                
  showstringspaces=false,
  showtabs=false,                  
  tabsize=2
}

\usepackage{listings}

\begin{document}

\title{AlphaX: An AI-Based Value Investing Strategy for the Brazilian Stock Market}
\author{Paulo Andr\'e Lima de Castro \\
\textit{Artificial Intelligence Applied to Finance Research Group- AIAF }\\
\textit{https://www.comp.ita.br/labsca/aiafGroup/}\\
Aeronautics Institute of Technology -ITA  \\
S\~ao Jos\'e dos Campos-SP, Brazil \\
pauloac@ita.br}
%}
%\institute{}
\maketitle

\begin{abstract}
Autonomous trading strategies have been a subject of research within the field of artificial intelligence (AI) for a considerable period. Various AI techniques have been explored to develop autonomous agents capable of trading financial assets. These approaches encompass traditional methods such as neural networks, fuzzy logic, and reinforcement learning, as well as more recent advancements, including deep neural networks and deep reinforcement learning. Many developers report success in creating strategies that exhibit strong performance during simulations using historical price data, a process commonly referred to as backtesting. However, when these strategies are deployed in real markets, their performance often deteriorates, particularly in terms of risk-adjusted returns. In this study, we propose an AI-based strategy inspired by a classical investment paradigm: Value Investing. Financial AI models are highly susceptible to look-ahead bias and other forms of bias that can significantly inflate performance in backtesting compared to live trading conditions. To address this issue, we conducted a series of computational simulations while controlling for these biases, thereby reducing the risk of overfitting. Our results indicate that the proposed approach outperforms major Brazilian market benchmarks. Moreover, the strategy, named AlphaX, demonstrated superior performance relative to widely used technical indicators such as the Relative Strength Index (RSI) and Money Flow Index (MFI), with statistically significant results. Finally, we discuss several open challenges and highlight emerging technologies in qualitative analysis that may contribute to the development of a comprehensive AI-based Value Investing framework in the future.
\end{abstract}

\section{Introduction}
\label{sec:intro}

The evolution of artificial intelligence (AI) in finance represents a gradual yet transformative integration of computational technologies into financial decision-making processes. This trajectory began with foundational theoretical developments in the 1950s and has progressed toward highly adaptive systems that significantly influence contemporary financial markets. Initial experimentation during the 1980s focused on expert systems, which subsequently gave way to the adoption of statistical models and algorithmic trading in the 1990s, marking the emergence of quantitative finance. In the 2000s, the proliferation of digital data coupled with advances in computing power enabled more sophisticated AI applications, such as high-frequency trading and fraud detection. The subsequent decade witnessed the mainstream integration of machine learning and natural language processing, exemplified by the rise of robo-advisors and sentiment analysis tools, thereby solidifying AI as a core component of modern financial infrastructure.

Nowadadys, AI is deeply embedded across the financial industry. Banks, asset managers and other Financial Institution are using AI for supporting activities such as
customer profiling, fraud detection, portfolio optimization, alpha seeking, market forecasting, compliance and customer service among other activities. There are many studies focused on fundamentalist and technical data~\cite{Katz:00, Castro:07,Johnson:10, Castro:10}, some work started to use alternative data such as news sentiment and textual analyses as data input for AI models~\cite{Prado:18}.

\subsection{Quant and AI Investing}

Quantitative investing, often called quant investing, is an investment approach that relies on mathematical models, algorithms, and data analysis to make decisions rather than human intuition or traditional subjective analysis. Instead of evaluating individual companies by reading reports or meeting with management, quant investors use historical data and statistical patterns to identify trading opportunities. These models systematically process vast amounts of information and execute trades based on predetermined rules. 

The process begins with generating an investment idea or hypothesis—for example, that stocks with low price-to-earnings ratios tend to outperform or that momentum can predict short-term gains. Once an idea is formed, investors collect and clean large datasets, which may include price histories, fundamental financial data, alternative data such as satellite images or transaction records, and sentiment from news or social media. This data is then transformed into meaningful features that a model can use.

Next, investors build statistical or machine learning models designed to detect patterns, forecast returns, or manage risk. These models might range from simple regressions to complex neural networks tailored to financial markets. After development, the strategy undergoes backtesting—running the model on historical data to evaluate its performance and robustness. If the strategy proves effective, it is deployed in live markets, often using automated systems that can trade thousands of assets simultaneously and rapidly.

Quant professionals have been using computational tools and AI techniques, as well as mathematical modeling, since the beginning of their work. It is not possible to really distinguish between Quant professionals and financial AI professionals today. They try to identify patterns in data, any type of data that is relevant, that helps them determine what will happen next and exploit this predictive power to make profitable trades. One of the most successful Quant companies, Renaissance Technologies and its founder Jim Simons are often referred to as pioneers of Quant investment in real markets~\cite{Zuckerman:19}. After its begin in 1982 and their success many other companies and hedge funds started to adopt Quant strategies at least as part of their decision Making. Nowadays, such methods are widely used by hedge funds such as and Two Sigma, Citadel and even retail platforms offer robo-advisory services~\cite{Wu:25,Zuckerman:19}.

\subsection{Strategies and Vanishing Alpha}

Strategies that aim to generate high returns by skillfully timing the selection or sizing of various asset positions~\cite{Narang:13} are called Alpha strategies. In contrast, Beta strategies replicate or slightly improve on the performance of a reference benchmark, such as the S\&P 500 or Brazil’s Ibovespa index.

When a particular market pattern is discovered, regardless how it was discovered, if more and more investors trade using it, they make price adjustments happen faster, eliminating the edge generated by it. Simply spreading a known pattern may cause it to disappear. Such phenomena is called \textbf{Vanishing alpha}.  Vanishing alphas makes quant firms very concerned about the confidentiality of their findings, models and even the technologies used by them.

In fact, Zuckerman starts his book about Jim Simon, \textit{The Man who solved the market}~\cite{Zuckerman:19},  describing how difficult was to get any information about his company. Employees and even competitors were reluctant to talk about it. The techniques and alpha findings are protected by an  'iron-clad, thirty-page nondisclosure agreements the firm forced employees to sign, preventing even retirees from divulging much'~\cite{Zuckerman:19}. This culture of secrecy is widespread within the Financial AI and quantitative investment community. It is interesting observe to note that~\textbf{Value Investing}, a different investment approach, the secrecy concerns are not that strong. We address it in the next section.

\section{Value Investing}

Value investing relies on deep fundamental analysis, looking for stocks priced below their true worth based on company financial and long-term prospects. Value investing can also suffer from the problem of vanishing alpha, though in a somewhat different way than purely quantitative strategies. People often overvalue short-term results over long-term outcomes because of human nature. In fact, a well known fact is that people are usually much more concerned
with now as opposed to later, which is called \textbf{present bias} in Behavioral Finance~\cite{Thaler:08}. 

Value investing and quantitative investing can both be applied across various investment horizons. However, quantitative investing is typically associated with short-term strategies, ranging from several months to just hours, minutes, or even milliseconds in the case of high-frequency trading (HFT). In contrast, value investing is closely tied to long-term positions, often held for several years. This does not mean that value investing cannot be used for short-term trades, or that quantitative investing cannot be applied to long-term strategies — both can be, and sometimes are. Nevertheless, they are more commonly employed in these traditional timeframes. 

Due to its long-term orientation and the tendency to counteract present bias, value investing may be more resilient to vanishing alpha. As Warren Buffett, a very famous value investor, once said when asked why more people do not follow such approach: “because no one wants to get rich slowly”.

The idea behind Value Investing is, in fact, quite simple: assesses a company's intrinsic worth by analyzing its financial statements, earnings potential, asset base, debt levels, competitive position, and overall business model. Then invest in companies that are trading below their estimated value. The goal is to obtain a comfortable \textbf{margin of safety} by purchasing undervalued assets. Value investors must be willing to go against prevailing market sentiment, often buying when many others are selling, based on the belief that the market will eventually correct itself and reflect the company’s actual worth.

Pioneered by Benjamin Graham, value investing requires processing a significant amount of information to form a deep understanding of the fundamentals of a company and a long-term approach~\cite{Graham:06}. In order to adopt a value investing approach, one needs a combination of financial knowledge, analytical  thinking and a long-term mindset. One needs to know \textbf{fundamental analysis} and define a set of companies that could be invested. This means to collect data from these target companies and being able to read and interpret the companies’ \textbf{financial statements} — including their income statement, balance sheet, and cash flow statement and understand key financial ratios such as the price-to-earnings (P/E) ratio, price-to-book (P/B) ratio, return on equity (ROE), and debt-to-equity ratio, among others. Equally important is the ability to estimate a company’s \textbf{intrinsic value}, which in Machine learning would be called a Regression problem. Then, an investor needs to select among the target companies those that provide a reasonable margin of safety, by only investing in those trading significantly below their intrinsic value. 

One must assess the quality of a company’s management, the durability of its competitive advantages, the stability of its business model, and the broader industry and regulatory environment. These assessments require to explore a company’s nonnumerical data, helping it understand its value or prospects.  Successful value investing also requires a good understanding of market psychology and its behavioral biases. Markets are often driven by emotion, with investors overreacting to short-term news, fear, or greed. These subjective and non-quantifiable aspects are usually refereed as \textbf{qualitative analysis}. It is is essential, but likely it is also the most challenging tasks regarding value investment. It is critical to acknowledge that Value investing also requires discipline to continue to and a long-term investment horizon~\cite{Fisher:96}. In the following section, we propose an AI strategy that automates these tasks, except for qualitative analysis, which we intend to address in section~\ref{sec:towards_qualitative}.

\section{AlphaX: An AI Strategy based on Value Investing}

AlphaX is an autonomous equity investment strategy designed for the Brazilian Stock Market. Its primary objective is to achieve returns that exceed the two main benchmarks of the Brazilian market—Selic (closely aligned with CDI which is Brazil’s benchmark interest rate) and Ibovespa~\cite{Assaf:14} (main stock market index in Brazil)—while maintaining a controlled risk profile. The strategy automates concepts and practices of Value Investing~\cite{Graham:06} and integrates fundamental and market data, thereby combining elements of fundamental and technical analysis through the application of advanced Artificial Intelligence techniques~\cite{Povoa:12,Appel:05}. Its development observed best practices to minimize the risk of model overfitting and enhance generalization~\cite{Luo:14,Prado:18}.

AlphaX constructs a portfolio by selecting risk assets based on its models and allocating capital accordingly. In the absence of suitable assets, capital is allocated to Treasury Selic instruments. The strategy incorporates a triple barrier framework—consisting of take-profit, stop-loss, and time-based constraints—where positions are closed at the end of the operational quarter, upon reaching the predefined target price, or following a price decline equal to or greater than 10%.

Risk mitigation is done in the selection of target companies, not through diversification as used in Markowitz Mean-Variance model~\cite{Markowitz:52}. We believe that it is only suitable for large capitals and for those who do not have good information about the businesses invested. AlphaX selects companies with low debt, good precification, and good profitability among all the analyzed companies. 

The investment horizon of the strategy is at least three months, but the use for longer periods favors better performance. Recommendations are made per quarter, since companies’ balance sheet data is released with this frequency. However, it is recommended that the investor aim for longer periods, making adjustments to the portfolio quarterly if needed.

\subsection{Data sources for AlphaX}

AlphaX gathers data from two sources: B3 (Brazilian Stock Exchange) and  CVM (CVM stands for Comissão de Valores Mobiliários, which is the Brazilian Securities and Exchange Commission (SEC) equivalent). 
It gets financial statements data about the target companies from CVM including Revenue, Operating Expenses, Gross Profit, EBIT, Net Income, Assets, Liabilities, Equity, Operation Cash Flow and so on.

Market data comes from B3 which is composed by open, maximum, minimum and close daily prices. The target assets were selected from stock traded on B3. In order to be selected, a company must have a long historical track record (a minimum of five years) and high liquidity (listed on B3’s main index). This enables more reliable projections. Sectors with high levels of indebtedness (such as retail) and financial institutions are not selected due to their particular risk characteristics. The complete list of target assets can be accessed seen in table~\ref{tab:assetsSX}.

\subsection{AlphaX Algorithm}

AlphaX computes four indicators: Profitability, Solvency, Valuation and Growth based on the fundamental and market data described in the previous section. These indicators are normalized from 1 (worst) to 5 (best). Then using those indicators and a set of Multiples commonly used in finance, such as Price-to-Earnings, Price-to-Book, Price-to-Sales and others, AlphaX makes a \textbf{price regression} based on an ensemble of Regressor algorithms (Random Forest and Regression to the mean).  

AlphaX selects for possible investments those companies with indicators above median, except for the Growth indicator which has a lower threshold. The selected assets are then ranked by  expected return - calculated as the percentage difference between projected price and current market price. The capital is allocated uniformly among the X best ranked assets indicated by the AlphaX strategy, where is X a parameter of the strategy indicating the maximum number of selectable assets. If there are less than X assets selected, the capital is allocated in the selected assets only also in a uniform way. In the extreme case that no asset is selected, the capital is allocated in a Selic bond, which may be interpreted as a risk-free asset in Brazilian market. 

The strategy also incorporates a \textbf{Triple Barrier} approach as proposed by Prado~\cite{Prado:18}. This method labels each investment position with three key barriers: Upper Barrier or \textbf{Take Profit} - if the asset reaches the projected price, the position in closed with profit; Lower Barrier or \textbf{Stop Loss} - if the asset reaches a predefined percentage below the entry price, the position is closed with loss and Vertical Barrier or \textbf{Maximum investment horizon} - a time limit after which the trade is closed if neither of the above barriers is hit. This limit is the date of the release of new financial statements. At this moment, AlphaX calculates again projected prices and select possible new assets. If the new selected assets are different to the previous ones, the position is closed with profit or loss. Wether the take profit or stop loss barrier are reached before the vertical barrier, the available capital is allocated in Selic bond. In the next section, we present simulated operations (backtesting) of AlphaX and compare them with benchmarks and some other autonomous trading strategies.

\section{Experiments and Results}

The AlphaX strategy was tested in simulated scenarios with real B3 quote data and balance sheet data from asset issuing companies, made available by the CVM. The testing period includes data from Feb, 2021 to May, 2025, which is equivalent to 18 quarters. The Selic rate data were obtained from the Central Bank website. It should also be considered that position change operations are made on the day of disclosure at the average price of the day (average of the maximum and minimum of the day). In our tests, we set X to 4, so up to four assets may be selected to investment. In order to avoid the look-ahead bias, we consider that the financial statements~\cite{Povoa:12} are available two months after the end of the quarter and not at the end of the quarter. Therefore, data is available on February 28 (relative to 4th quarter), May 31 (1st quarter), August 31 (2nd quarter) and November 30 (3rd quarter).  A total of 36 common or preferred shares were selected. The complete list is presented in Table~\ref{tab:assetsSX}.

\begin{table} [ht]%[htbp,width=0.9\textwidth]
	\begin{center}
		\begin{tabular}{|| l |l | l |l | l|| }
		 \hline	    & Ibovespa & Selic  & NIbov  & AlphaX  \\
\hline   Total Return (\%) & 16.6 & 55.1 & 72.6  & \BlackCell{97.9}  \\
\hline    CAGR  (\%) &  3.6  & 6.2 & 13.3 &  \BlackCell{16.9} \\
        \hline
		\end{tabular}
	\end{center}
	\caption{Results for AlphaX vs Indexes.A black cell indicates the best performance. CAGR stands for Compound Annual Growth Rate.}
	\label{tab:ResultsDailyAlphaX}
\end{table}

In the simulated results, the AlphaX strategy outperformed the Ibovespa and Selic regarding Total Return and CAGR (Compound Annual Growth Rate) as presented in Table~\ref{tab:ResultsDailyAlphaX}. We also tested a normalized version of Ibovespa, called NIbov in Table~\ref{tab:ResultsDailyAlphaX}, in order to control for possible Survivorship bias as described in~\cite{Luo:14}. It refers to the problem of selecting  well-performing companies while ignoring the others. NIbov is calculated with weights of the selected companies for AlphaX using normalized Ibovespa´s weights. AlphaX also overcomed NIbov presenting 97.9\% of Total return against 72.6\%. The next section provides detailed information about the simulations.

\begin{table}[ht] %[htbp,width=0.9\textwidth]
\begin{center}
\begin{tabular}{|| c| c| c ||}
\hline \# & Quarter  & Invested Stocks \\
\hline 1  & Q4 2020 & BRKM5 \\ % February 28, 2021
\hline 2  &  Q1 2021 & ABEV3,ENEV3,TAEE11,USIM5 \\ % May 31, 2021
\hline 3  &   Q2 2021 & USIM5, GGBR4, GOAU4 \\ % August 31, 2021
\hline 4  & Q3 2021    & GGBR4,GOAU4, CSNA3, PETR3 \\ %  November 30, 2021
\hline 5  & Q4 2021 & GOAU4, KLBN11, WEGE3, CSAN3
 \\
\hline 6  & Q1 2022  & GGBR4, GOAU4, PRIO3, SUZB3 \\
\hline 7  & Q2 2022  & GOAU4, PETR3, CSNA3, SUZB3
\\
\hline 8  & Q3 2022   & PRIO3, PETR3, GOAU4
 \\
\hline 9  & Q4 2022 & No stocks \\
\hline 10  & Q1 2023  & CMIG4, SUZB3
 \\
\hline 11  & Q2 2023  & SLCE3, GGBR4, SUZB3, GOAU4
\\
\hline 12  & Q3 2023   & CMIG4, GGBR4
 \\
\hline 13  & Q4 2023 & CMIG4, UGPA3
 \\
\hline 14  & Q1 2024  & SUZB3, PRIO3, WEGE3, TAEE11
 \\
\hline 15  & Q2 2024  & GOAU4, GGBR4, ABEV3
\\
\hline 16  & Q3 2024  & PRIO3
 \\
\hline 17  & Q4 2024 &  ELET3
\\
\hline 18  & Q1 2025  &  ABEV3
\\

\hline 
\end{tabular}
\end{center}
\caption{AlphaX´s Invested Stocks per Quarter}
\label{tab:allocations}
\end{table}

Simulations of the 18 quarters were carried out with special care to mitigate the risk of overfitting, following the techniques indicated by~\cite{Luo:14}. In particular, considering the entry point of each position as the average between the maximum and minimum prices of the day, to make the entry price more realistic and not biased. The assets selected for investment by AlphaX in each quarter are presented in Table~\ref{tab:allocations}

\subsection{AlphaX versus Technical Strategies}

\begin{table} [ht]%[htbp,width=0.9\textwidth]
	\begin{center}
		\begin{tabular}{|| l |l | l |l | l  | l|| }
		 \hline	      & RSI &  Stochastic & MFI    & AlphaX  \\
\hline   Total Return (\%) & 19.6 & 22.8 &  44.1 &  \BlackCell{97.9}  \\
\hline   CAGR (\%) &  4.2 & 4.8  & 8.7 &  \BlackCell{16.9} \\

\hline   Annualized Sharpe Ratio& 0.29 & 0.35 &  0.43 & \BlackCell{0.98} \\
\hline   Annualized Sortino &0.51 &  0.58 &   0.72 & \BlackCell{1.51} \\
\hline   Max Drawdown (\%) &  -45.9 & -23.0 & -40.4 &  \BlackCell{-20.8} \\

        \hline
		\end{tabular}
	\end{center}
	\caption{Results for AlphaX and some Technical Strategies. A black cell indicates the best performance. CAGR stands for Compound Annual Growth Rate.}
	\label{tab:ResultsStrategies}
\end{table}

We compared AlphaX to three well known technical strategies: RSI (Relative Strength Index)~\cite{Appel:05}, Stochastic~\cite{Katz:00} and MFI (Money Flow Indicator)~\cite{Appel:05}. 
The Relative Strength Index (RSI) is a momentum oscillator that measures the speed and magnitude of recent price movements. It ranges from 0 to 100 and helps identify overbought (above 70) or oversold (below 30) conditions in the target asset. The Stochastic Oscillator compares an asset’s closing price to its price range over a specific period (in our setup 14 days). It tries to identify overbought and oversold conditions and ranges above zero and 100, if above 80 it indicates a sell signal, and below 20 it is a buy signal.
The Money Flow Index (MFI) measures the flow of money into and out of an asset over a given period, in our setup 14 days. MFI incorporates price and volume, making it a volume-weighted momentum oscillator. It ranges from 0 to 100, where values above 80 suggest an overbought condition and values below 20 indicate an oversold condition. 

AlphaX has shown superior performance compared to all of them in terms of Return (CAGR, Total Return), Risk (Max Drawdown) and Risk-adjusted returns (An. Sharpe Ratio and An. Sortino), as presented in Table~\ref{tab:ResultsStrategies}.

\subsection{Probabilistic Sharpe Ratio}

 Portfolio managers are often evaluated based on Sharpe Ratio, which measures the risk-adjusted return of an investment compared to a risk-free asset. It helps an investor to assess whether the returns generated by a portfolio are due to a skillful management or simply the result of taking on more risk. 
 
 Despite its wide use, Sharpe ratio has several limitations~\cite{Bailey:12}. First, it assumes that returns follow a normal distribution, which often doesn't align with real-world market behavior, especially for assets with returns that exhibit skewness or fat tails. Additionally, the Sharpe ratio relies solely on volatility (standard deviation) as a measure of risk, ignoring other crucial risk factors such as liquidity risk, drawdown risk, or the potential for large, sudden losses. Finally, it fails to account for tail risks—extreme market events, like "black swan" occurrences. These limitations make the Sharpe ratio useful but incomplete when evaluating the risk-adjusted performance of a portfolio. 
 
 Bailey and Prado~\cite{Bailey:12} proposed a metric called Probabilistic Sharpe Ratio or PSR. It adjusts the traditional Sharpe ratio to account for non-normal return distributions and limited sample sizes. It estimates the probability that a strategy's Sharpe ratio is higher than a chosen threshold with a level of confidence. In order to evaluate AlphaX Strategy, we used an open source implementation of PSR~\cite{Castro:21}.

\begin{table} [ht]%[htbp,width=0.9\textwidth]
	\begin{center}
		\begin{tabular}{|| l |l | l | l | l | l|| }
		 \hline	      & RSI &  Stochastic & MFI    & AlphaX  \\

\hline   PSR(0)& 52.3\% & 89.8\% & 81.4\%& \BlackCell{97.4\%} \\
\hline    minTRL(0)& 338.6 & 7.2 & 14.6 & 3.1  \\
\hline   PSR(0.01)& 39.3\% & 82.6\% & 71.3\% & \BlackCell{94.8 \%}  \\
\hline    minTRL(0.01)&158.9 &13.3 & 36.8 & 4.4 \\
\hline   PSR(0.1)&0.1\%& 2.1\% &0.8\% & 11.6\%  \\
\hline    minTRL(0.1)&1.1 &2.8 &2.0 & 8.2 \\
        \hline
		\end{tabular}
	\end{center}
	\caption{Probabilistic Sharpe Ratio for for AlphaX and technical trading strategies. A black cell indicates that the strategy overcame the proposed Sharpe ratio threshold considering the minimum track record (minTRL).}
	\label{tab:Results3}
\end{table}

In Table~\ref{tab:Results3}, the results show that AlphaX has an estimated Sharpe Ratio above zero, PSR(0), with 97.4\% level of confidence and above 0.01 with 95\% level of confidence. None of the analyzed technical strategies (RSI, Stochastic and MFI) presented such performance. MFI presented the second best performance with an estimated Sharpe ratio above zero with just 81\% level of confidence. No technical strategy reaches a 95\% level of confidence. Bailey and Prado applied PSR to 33 HFR (Hedge Fund Research) indices. HFR indices are real data from January 1st 2000 to May 1st 2011. It consists of 134 monthly observations and using a confidence level of 95\%. 
Observe that HFR data is composed by funds managed by human experts, not by autonomous trading strategies. 
Their results showed that overcome a 0.5 threshold is a hard task. Only 9 indices presented an annualized Probabilistic Sharpe Ratio of 0.5 with a 95\% confidence level, and some of them (4 in 33) did not presented an annualized Probabilistic Sharpe Ratio over zero.

\section{Towards Qualitative Analysis}
\label{sec:towards_qualitative}

In practical investment management performance, managers evaluators incorporate both quantitative and qualitative elements, with the latter usually receiving more attention than the former~\cite{Maginn:07}. We use the term \textbf{Qualitative Analysis} refer to such aspects. They include at least four complex tasks: analyzing textual information complementary to financial statements~\cite{Brown:12,Damodaran:10,Marks:11}, analyzing the competitiveness of a given company against its competitors~\cite{Blank:05,Kotler:14}, possible changes in regulatory environment~\cite{Mankiw:18} and market psychology evolution~\cite{Kahneman:13}, which includes aspects regarding behavioral biases. 

Behavioral biases are systematic errors in decision making that can lead investors to make irrational or suboptimal decisions~\cite{Kahneman:79,Kahneman:13,Thaler:08}. These biases often stem from psychological tendencies, emotions, or cognitive shortcuts, which cause investors to deviate from objective analysis and rational behavior~\cite{Castro:14,Castro:16A,Thaler:08}. For example, herding bias make investors follow the crowd even if it goes against their own analysis, and recency bias, where recent events are given too much weight in decision making. These biases can distort risk assessment, timing, and ultimately damaging making the market to price assets incorrectly. Some studies have been able to identify the occurence of biases like that in real markets~\cite{Khorana:00} including the Brazilian Market~\cite{Barbedo:21}.

Large Language Models (LLMs) represent a potentially transformative tool in Qualitative analysis. In the context of company and sector analysis, These models demonstrate a remarkable capacity to extract meaningful insights from large-scale, unstructured textual data originating from diverse sources in real time~\cite{Chen:23,Finllama:24}. Moreover, LLMs show significant promise in identifying and interpreting potential shifts in the regulatory landscape, thereby enhancing the monitoring and analysis of evolving market conditions. Transformers, the base technology behind LLM, may also have significant impact in AI investing, specially in terms of Alpha seeking and risk management~\cite{Castro:24}.

\begin{table}[ht] %[htbp,width=0.9\textwidth]
\begin{center}

\begin{tabular}{|| c| c| c ||}
%\hline \multicolumn{2}{|c|}{}\\
\hline \# & Asset ID  & Company \\
\hline 1  & ALPA4  & ALPARGATAS \\
\hline 2  & ABEV3  & AMBEV SA \\
\hline 3  & B3SA3  & B3 SA \\
\hline 4  & BRKM5  & BRASKEM \\
\hline 5  & BRFS3  & BRF S.A. \\
\hline 6  & CMIG4  & CEMIG \\
\hline 7  & COGN3  & KROTON EDUCACIONAL SA \\
\hline 8  & CSAN3  & COSAN SA INDUSTRIA E COMERCIO \\
\hline 9  & CPFE3  & CPFL ENERGIA SA \\
\hline 10  & DXCO3  & DURATEX SA \\
\hline 11  & ELET3  & ELETROBRAS \\
\hline 12  & EMBR3  & EMBRAER \\
\hline 13  & ENGI11  & ENERGISA \\
\hline 14  & ENEV3  & ENEVA SA \\
\hline 15  & EQTL3  & EQUATORIAL ENERGIA SA \\
\hline 16  & EZTC3  & EZ TEC EMPREEND. PAR SA \\
\hline 17  & FLRY3  & FLEURY SA \\
\hline 18  & GGBR4  & GERDAU S.A. \\
\hline 19  & GOAU4  & METALURGICA GERDAU HOLDING SA \\
\hline 20  & KLBN11  & KLABIN \\
\hline 21  & RENT3  & LOCALIZA RENT A CAR \\
\hline 22  & BEEF3  & MINERVA FOODS S/A \\
\hline 23  & PETR3  & PETROBRAS \\
\hline 24  & PRIO3  & PETRO RIO S.A. \\
\hline 25  & RAIL3  & ALL - AMÉRICA LATINA LOGÍSTICA SA \\
\hline 26  & CSNA3  & CSN \\
\hline 27  & SLCE3  & SLC AGRICOLA SA \\
\hline 28  & SUZB3  & SUZANO PAPEL E CELULOSE SA \\
\hline 29  & TAEE11  & TRANSMISSORA ALIANÇA E. ELÉT. \\
\hline 30  & VIVT3  & TELEFÔNICA BRASIL \\
\hline 31  & TOTS3  & TOTVS SA \\
\hline 32  & UGPA3  & ULTRAPAR PARTICIPAÇÕES SA \\
\hline 33  & USIM5  & USIMINAS \\
\hline 34  & WEGE3  & WEG SA \\
\hline 35  & YDUQ3  & ESTACIO PARTICIPAÇÕES SA \\
\hline 36  & MOTV3  & MOTIVA \\
\hline 
\end{tabular}
\end{center}
\caption{List of B3 Assets used in Simulations}
\label{tab:assetsSX}
\end{table}

\section{Conclusions}

In this study, we propose an autonomous trading strategy that integrates fundamental and market data to guide capital allocation. This strategy, called AlphaX, employs artificial intelligence techniques to automate the principles of value investing, with an emphasis on achieving consistent returns while maintaining controlled risk over investment horizons of three months or longer. The strategy was evaluated through computational simulations (backtesting) using real historical data spanning 18 quarters. Results indicate that AlphaX outperformed the Ibovespa and exceeded the Selic. Additionaly, AlphaX outperformed three well known technical strategies (RSI,Stochastic and MFI) in terms of Return, Risk and Risk-adjusted Return.

We made AlphaX’s projections publicly accessible via the platform www.finbrain.com.br. We also noted that developing and backtesting strategies based on fundamental data is significantly more challenging than relying solely on market data. To the best of our knowledge, all freely available backtesting tools are primarily designed for technical analysis and lack support for fundamental-based approaches. To address this gap, we plan to release the code developed for executing our backtests and implementing our strategy under an open-source license in the future.

Value investing traditionally involves subjective analyses of a company’s fundamentals, interpretative assessments of financial statements, and evaluations of managerial competence in sustaining competitive advantage. The strategy proposed in this paper does not encompass these qualitative aspects. Incorporating such capabilities into autonomous agents would require not only extracting information from natural language text but also reasoning about company fundamentals across diverse market sectors. Recent advances in Transformer architectures and Large Language Models (LLMs) have begun to revolutionize financial applications~\cite{li:23}. Nevertheless, significant challenges remain in developing autonomous strategies capable of performing real~\textbf{qualitative analysis}. We discussed these challenges in Section~\ref{sec:towards_qualitative} and possible directions for addressing them.

\textbf{DISCLAIMER}:\textit{This text is for informational purposes only and does NOT constitute an analysis report. The input data for the models are collected from various sources such as CVM, B3, BACEN, among others. However, there may be errors in their processing; we are not responsible for any errors. Some of the information is generated by Machine Learning/Artificial Intelligence models, but there is no guarantee that it will be correct. This document is not an investment recommendation.}

\bibliographystyle{unsrt}
\bibliography{bibliopa}

\begin{thebibliography}{10}

\bibitem{Katz:00}
Jefferey Katz and Donna McCormick.
\newblock {\em The Encyclopedia of trading strategies}.
\newblock McGraw-Hill, New York, 2000.

\bibitem{Castro:07}
Paulo~Andre Castro and Jaime~Simao Sichman.
\newblock Towards cooperation among competitive trader agents.
\newblock In {\em Proceedings of 9th ICEIS.}, pages 138--143, Funchal, Portugal, 2007.

\bibitem{Johnson:10}
Barry Johnson.
\newblock {\em Algorithmic Trading and Direct Market Access}.
\newblock Myeloma Press, London, 2010.

\bibitem{Castro:10}
Carlos H.~Dejavite Araujo and Paulo Andre~L. Castro.
\newblock Towards automated trading based on fundamentalist and technical data.
\newblock In {\em Proceedings of 20 th SBIA. LNAI}, pages 704--715, Sao Bernado do Campo, Brazil, 2010. Springer-Verlag.

\bibitem{Prado:18}
Marcos~Lopez de~Prado.
\newblock {\em Advances in Financial Machine Learning.}
\newblock Wiley, New York, 2018.

\bibitem{Zuckerman:19}
Gregory Zuckerman.
\newblock {\em The Man Who Solved the Market: How Jim Simons launched the quant revolution}.
\newblock Portfolio Penguim, 2019.

\bibitem{Wu:25}
Zijing Wu.
\newblock Baiont’s feng ji: Quant managers who don’t adopt ai will be eliminated by the market, 2025.

\bibitem{Narang:13}
Rishi~K. Narang.
\newblock {\em Inside the black box: a simple guide to quantitative and high-frequency trading}.
\newblock Wiley finance series. John Wiley \& Sons, Inc, Hoboken, New Jersey, second edition edition, 2013.

\bibitem{Thaler:08}
Richard~H. Thaler and Cass~R. Sunstein.
\newblock {\em Nudge}.
\newblock Yale University Press, New Haven, CT and London, 2008.

\bibitem{Graham:06}
Benjamin Graham.
\newblock {\em The Intelligent Investor}.
\newblock Harper Business, Revised Edition, New York, 2006.

\bibitem{Fisher:96}
Philip~A. Fisher.
\newblock {\em Common Stocks and Uncommon Profits and Other Writings}.
\newblock John Wiley and Sons, 1996.

\bibitem{Assaf:14}
Alexandre~Assaf Neto.
\newblock {\em Mercado Financeiro. 12a. Ed.}
\newblock Ed. Atlas, São Paulo, 2014.

\bibitem{Povoa:12}
Alexandre Póvoa.
\newblock {\em Valuation: Como precificar a?es}.
\newblock Campus-Elsevier, Rio de Janeiro, 2012.

\bibitem{Appel:05}
Gerald Appel.
\newblock {\em Technical Analysis: Power Tools for Active Investors}.
\newblock Financial Times Prentice Hall, New York, 2005.

\bibitem{Luo:14}
Y.~Luo, M.~Alvarez, J.~Jussa S.~Wang, A.~Wang, and G.~Rohal.
\newblock Seven sins of quantitative investing, September 8 2014.

\bibitem{Markowitz:52}
Harry~M. Markowitz.
\newblock Portfolio selection.
\newblock {\em Journal of Finance}, 7(1):77--91, 1952.

\bibitem{Bailey:12}
David~H. Bailey and Lopez de~Prado.
\newblock The sharpe ratio efficient frontier.
\newblock {\em Journal of Risk}, 15:3--44, Feb 2012.

\bibitem{Castro:21}
Murilo~Sibrao Bernardini and Paulo André~Lima de~Castro.
\newblock Is it a great autonomous fx trading strategy or you are just fooling yourself?
\newblock {\em ArXIV}, 2021.

\bibitem{Maginn:07}
John~L. Maginn.
\newblock {\em Managing investment portfolios : a dynamic process}.
\newblock CFA Institute. John Wiley and Sons, Inc., Hoboken, New Jersey, third edition. edition, 2007.

\bibitem{Brown:12}
Frank~K. Reilly and Keith~C. Brown.
\newblock {\em Investment Analysis \& Portfolio Management}.
\newblock South-Western, Cengage Learning, New York, 2012.

\bibitem{Damodaran:10}
Aswath Damodaran.
\newblock {\em Applied Corporate Finance}.
\newblock Wiley, New York, NY, 2010.

\bibitem{Marks:11}
Howard Marks.
\newblock {\em The most important thing: uncommon sense for the thoughtful investor}.
\newblock Columbia Press, New York, 2011.

\bibitem{Blank:05}
Steve~Gary Blank.
\newblock {\em The Four Steps to the Epiphany: Successful Strategies for Products That Win}.
\newblock K \& S Ranch, New York, 2005.

\bibitem{Kotler:14}
Philip Kotler and Gary Armstrong.
\newblock {\em Principles of Marketing}.
\newblock Prentice-Hall, 15nd edition, 2014.

\bibitem{Mankiw:18}
N.~Gregory Mankiw.
\newblock {\em Principles of Economics. 8th Edition}.
\newblock Cengage Learning, New York, 2018.

\bibitem{Kahneman:13}
Daniel Kahneman.
\newblock {\em Thinking, Fast and Slow}.
\newblock Farrar Straus Giroux, New York, 2013.

\bibitem{Kahneman:79}
Daniel Kahneman and Amos Tversky.
\newblock Prospect theory: An analysis of decision under risk.
\newblock {\em Econometrica}, 47(2):263--291, March 1979.

\bibitem{Castro:14}
Paulo Andre~Lima Castro and Simon Parsons.
\newblock Modeling agent's preferences based on prospect theory.
\newblock In {\em Proccedings of the 8th Multidisciplinary Workshop on Advance in Preference Handling (MPREF 2014)}, pages 32--36, Quebec, Canada, July 2014. co-located with AAAI-2014.

\bibitem{Castro:16A}
Paulo André~Lima de~Castro, Anderson Rodrigo~Barreto Teodoro, Luciano~Irineu de~Castro, and Simon Parsons.
\newblock Expected utility or prospect theory: Which better fits agent-based modeling of markets?
\newblock {\em Journal of Computational Science}, 17:97 -- 102, 2016.

\bibitem{Khorana:00}
Eric~C Chang, Joseph~W Cheng, and Ajay Khorana.
\newblock An examination of herd behavior in equity markets: An international perspective.
\newblock {\em Journal of Banking and Finance}, 24(10):1651--1679, 2000.

\bibitem{Barbedo:21}
E.~Camilo-da-Silva P.~Signorelli and C.H. Barbedo.
\newblock n examination of herding behavior in the brazilian equity market.
\newblock {\em Brazilian Business Review}, 18:236–254, 2021.

\bibitem{Chen:23}
Yinheng Li, Shaofei Wang, Han Ding, and Hang Chen.
\newblock Large language models in finance: A survey.
\newblock In {\em Proceedings of the Fourth ACM International Conference on AI in Finance}, ICAIF '23, page 374–382, New York, NY, USA, 2023. Association for Computing Machinery.

\bibitem{Finllama:24}
Thanos Konstantinidis, Giorgos Iacovides, Mingxue Xu, Tony~G. Constantinides, and Danilo Mandic.
\newblock Finllama: Financial sentiment classification for algorithmic trading applications, 2024.

\bibitem{Castro:24}
Lucas Coelho~e Silva, Gustavo de~Freitas Fonseca, and Paulo Andre~L. Castro.
\newblock Transformers and attention-based networks in quantitative trading: a comprehensive survey.
\newblock In {\em Proceedings of the 5th ACM International Conference on AI in Finance}, ICAIF '24, page 822–830, New York, NY, USA, 2024. Association for Computing Machinery.

\bibitem{li:23}
Yinheng Li, Shaofei Wang, Han Ding, and Hang Chen.
\newblock Large {Language} {Models} in {Finance}: {A} {Survey}.
\newblock In {\em 4th {ACM} {International} {Conference} on {AI} in {Finance}}, pages 374--382, Brooklyn NY USA, November 2023. ACM.

\end{thebibliography}

\end{document}